\documentclass[sigconf,nonacm]{acmart}
\renewcommand\footnotetextcopyrightpermission[1]{}
\usepackage{graphicx}
\graphicspath{{figures/}{./}}
\usepackage{amsmath}
\usepackage{booktabs}
\usepackage{multirow}
\usepackage{threeparttable}
\usepackage{array}
\usepackage{tabularx}
\usepackage{xcolor}
\usepackage{enumitem}
\usepackage{dblfloatfix}
\usepackage{placeins}
\usepackage{eso-pic}
\newcommand{\dataset}{\textsc{ONOTE}}
\newcommand{\method}{\textsc{ONOTE}}
\newcommand{\gain}[1]{\textcolor{green!45!black}{\tiny(+#1)}}
\setcopyright{none}
\acmDOI{}
\acmISBN{}
\acmYear{2027}
\copyrightyear{2027}

\title{ONOTE: Hypergraph-Grounded Omnimodal Reasoning for Computational Music Science}

\author{Menghe Ma}
\authornote{Equal contribution.}
\affiliation{%
  \department{School of Digital Media and Design Arts}
  \institution{Beijing University of Posts and Telecommunications}
  \city{Beijing}
  \country{China}}
\email{mamenghe@bupt.edu.cn}

\author{Siqing Wei}
\authornotemark[1]
\affiliation{%
  \department{School of Future}
  \institution{Beijing University of Posts and Telecommunications}
  \city{Beijing}
  \country{China}}
\email{wsq@bupt.edu.cn}

\author{Yuecheng Xing}
\authornotemark[1]
\affiliation{%
  \department{School of Artificial Intelligence}
  \institution{Beijing University of Posts and Telecommunications}
  \city{Beijing}
  \country{China}}
\email{xingyuecheng@bupt.edu.cn}

\author{Ziyue Zhu}
\affiliation{%
  \department{State Key Laboratory of Networking and Switching Technology}
  \institution{Beijing University of Posts and Telecommunications}
  \city{Beijing}
  \country{China}}
\email{zyzhu@bupt.edu.cn}

\author{Zhenghong Lin}
\affiliation{%
  \department{Artificial Intelligence Innovation Center}
  \institution{China Unicom (Fujian) Industrial Internet Co., Ltd.}
  \city{Fuzhou}
  \country{China}}
\email{hongzhenglin970323@gmail.com}

\author{Yaheng Wang}
\affiliation{%
  \department{School of Future}
  \institution{Beijing University of Posts and Telecommunications}
  \city{Beijing}
  \country{China}}
\email{wangyaheng@bupt.edu.cn}

\author{Fanhong Meng}
\affiliation{%
  \department{Institute of Music Education}
  \institution{China Conservatory of Music}
  \city{Beijing}
  \country{China}}
\email{fanhong.meng@outlook.com}

\author{Peijun Han}
\affiliation{%
  \department{Institute of Music Education}
  \institution{China Conservatory of Music}
  \city{Beijing}
  \country{China}}
\email{peijunhan0517@163.com}

\author{Luu Anh Tuan}
\affiliation{%
  \department{College of Computing and Data Science}
  \institution{Nanyang Technological University}
  \city{Singapore}
  \country{Singapore}}
\email{anhtuan.luu@ntu.edu.sg}

\author{Haoran Luo}
\authornote{Corresponding author.}
\affiliation{%
  \department{College of Computing and Data Science}
  \institution{Nanyang Technological University}
  \city{Singapore}
  \country{Singapore}}
\email{haoran.luo@ntu.edu.sg}

\renewcommand{\shortauthors}{Ma et al.}
\begin{document}

\AddToShipoutPictureFG*{%
  \AtPageLowerLeft{%
    \hspace*{\dimexpr 1in+\oddsidemargin\relax}%
    \raisebox{0.95in}[0pt][0pt]{%
      \parbox[b]{\columnwidth}{%
        \footnotesize
        Data: \url{https://huggingface.co/datasets/Weisiqing123/ONOTE}\par
        Code: \url{https://github.com/T12knightally/ONOTE}%
      }%
    }%
  }%
}

\begin{abstract}
Omnimodal notation processing, centered on sheet music, is a controlled scientific setting in which auditory, visual, symbolic, and physical representations must encode the same musical events. Yet existing work remains fragmented across recognition and transcription, rarely testing structural consistency across notation systems. Western-staff bias and underspecified model judges further conceal errors in pitch, timing, ordering, and instrument-specific constraints.We introduce \dataset, a unified framework that treats music as a scientifically structured domain of measurable cross-representation correspondences. Its test-only benchmark draws on a diverse collection of musical sources covering staff, Jianpu, and tablature across varied genres, instruments, and structural conditions, with aligned multimodal derivatives. Four complementary tasks cover score understanding, notation conversion, audio transcription, and symbolic generation, testing pitch and duration ordering, output syntax, and disclosed instrument-specific constraints. \dataset{} also constructs a provenance-bearing proposition hypergraph from external music-theory materials for entity- and hyperedge-based evidence retrieval.Deterministic validity checks, disclosed structural-compliance SMG scoring, and controlled RAG comparisons reveal gaps between visual recognition and structure-preserving outputs. Results separate perception from music-theory application and structural or physical constraint satisfaction. \dataset{} provides an auditable framework for studying representation invariance and knowledge-grounded intervention in computational music science.
\end{abstract}



\maketitle

\section{Introduction}
\begin{figure}[t]
    \centering
    \captionsetup{belowskip=-2pt}
    \includegraphics[width=\columnwidth]{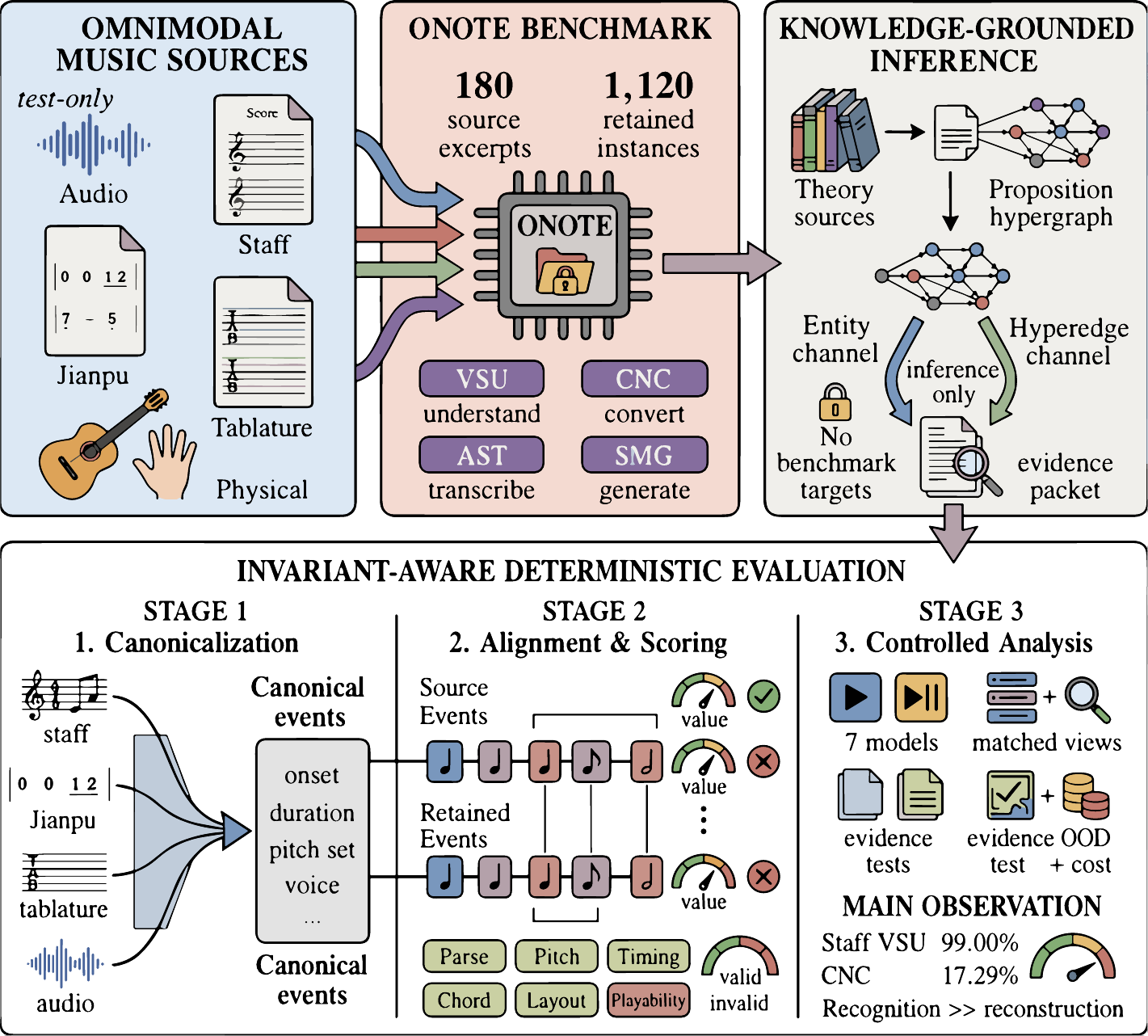}
    \caption{Scientific formulation of omnimodal notation reasoning. Audio, score, symbolic, and physical views map to a shared canonical event while evaluation checks pitch, time, ordering, and disclosed physical constraints.}
    \label{fig:scientific-formulation}
    \Description{Scientific task diagram connecting music representations, the ONOTE benchmark, knowledge-grounded inference, and invariant-aware deterministic evaluation.}
\end{figure}

Omnimodal large language models can process natural language, images, and audio, yet omnimodal notation processing remains a demanding scientific reasoning problem. Music requires precise spatial--temporal alignment across auditory observations, visual scores, symbolic languages, and instrument mechanics (Figure~\ref{fig:scientific-formulation}). Unlike ordinary text, each representation must preserve pitch, onset, duration, simultaneity, meter, and instrument-specific physical constraints. Advances in optical music recognition (OMR) \cite{calvozaragoza2023omr,calvozaragoza2018end}, automatic music transcription (AMT) \cite{hawthorne2018onsets,gardner2022mt3}, and symbolic generation \cite{lu2023musecoco,yuan2024chatmusician} have produced strong specialized systems and code-based representations. However, these lines of work remain fragmented: recognition, transcription, conversion, and generation are usually evaluated in isolation. They therefore do not reveal whether the same event structure survives a change of modality or notation. A model can recognize local symbols while failing to reconstruct the syntax, chord structure, or performance constraints of a complete score.

Current models and benchmarks also exhibit severe notation bias, prioritizing Western staff while underrepresenting globally used systems such as Jianpu and guitar tablature. Subjective ``LLM-as-a-judge'' evaluation compounds this problem: plausible-looking outputs can receive favorable scores despite wrong pitches, broken meter, lost simultaneity, or constraint-violating fingering.

This motivates event-level evaluation that tests representation changes directly rather than rewarding plausible surface forms.

To address these limitations, we introduce \dataset, a unified framework that combines a test-only benchmark covering three notation systems and four task tracks with retrieval over external music-theory propositions. Its deterministic pipeline canonicalizes events, while dual-channel hypergraph retrieval supplies provenance-bearing rules without accessing benchmark targets.

Experiments across seven omnimodal models reveal a clear discrepancy between multiple-choice accuracy and ordered-sequence conversion or transcription. Controlled Base--RAG comparisons estimate the effect of adding retrieved evidence, while a frozen end-to-end comparison evaluates the complete ONOTE and legacy retrieval configurations. Auditable case studies separate retrieval, structured-decoding, generation-diversity, and constraint-compliance failures. Using task-specific metrics, \dataset{} diagnoses modality-specific failures and retrieval gains in computational music.

\begin{figure*}[t]
\centering
\includegraphics[width=\textwidth]{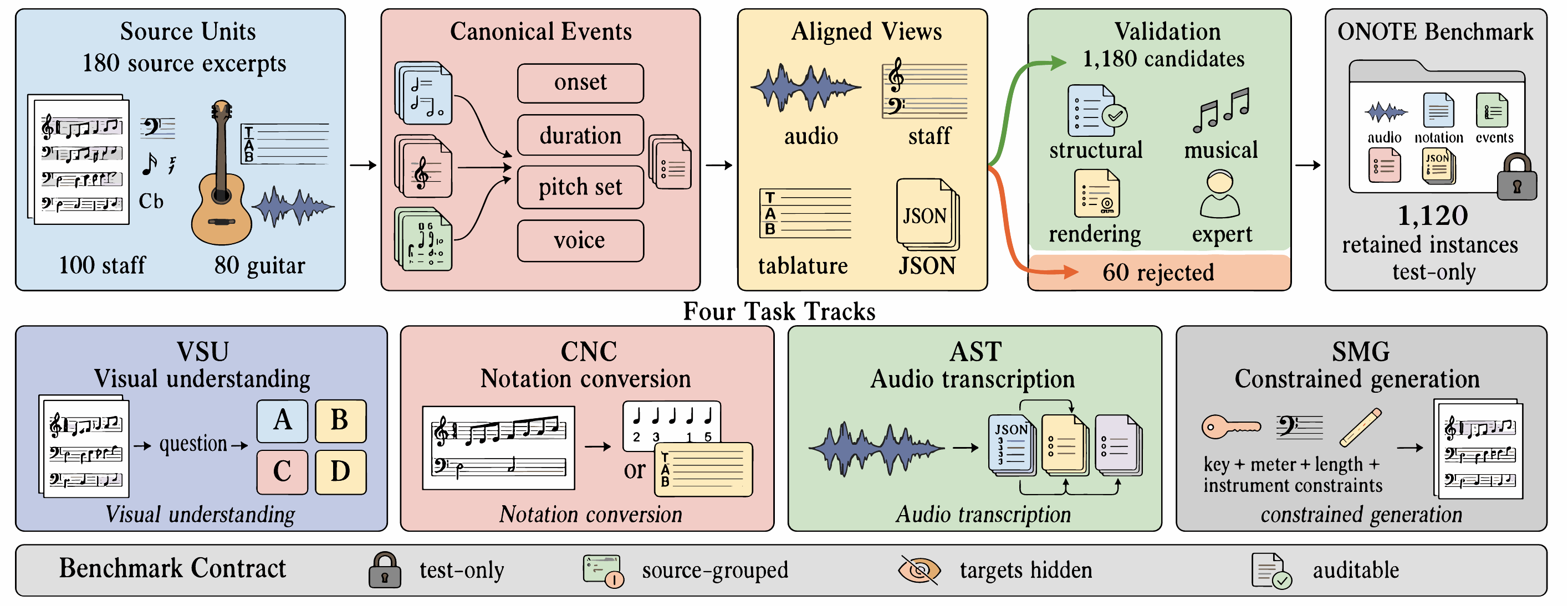}
\caption{End-to-end \dataset{} workflow. The test-only benchmark branch is isolated from the external-theory branch; retrieval enters only at model inference, after benchmark targets have been hidden.}
\label{fig:workflow}
\Description{ONOTE workflow with separate benchmark and external-knowledge branches feeding deterministic evaluation.}
\end{figure*}

\section{Related Work}
\subsection{Structured Multimodal Reasoning}
Structured multimodal reasoning combines heterogeneous observations and
domain knowledge under explicit constraints and provenance. Computational
music is a useful testbed because formal rules connect audio, notation,
symbolic events, and performance while permitting representation-specific
errors to be traced to observable event mismatches. These correspondences
also expose whether pitch, timing, ordering, and physical constraints remain
consistent across modalities. We test whether shared musical structure
survives cross-representation transformations using deterministic,
provenance-linked audits.

\subsection{Computational Music Understanding}
MAESTRO \cite{hawthorne2019maestro}, Slakh \cite{manilow2019slakh}, and GuitarSet \cite{xi2018guitarset} provide aligned or instrument-specific collections; MusiXQA \cite{chen2025musixqa} targets visual music understanding. MT3 \cite{gardner2022mt3}, Basic Pitch \cite{bittner2022basicpitch}, and tablature estimators \cite{wiggins2019tablature} separate general-model limitations from task difficulty. \dataset{} instead tests cross-representation preservation, including Jianpu.

MuseCoco \cite{lu2023musecoco}, ChatMusician \cite{yuan2024chatmusician}, and NotaGen \cite{wang2025notagen} show the value of music-specific symbolic languages. We separate deterministic renderability, meter, pitch, rhythm, and fingering-constraint compliance from aesthetic preference.

Recent art-and-AI research spans generative music, visual creation, human--AI co-creativity, design, cultural heritage, narrative practice, aesthetics, authorship, and copyright \cite{artai2025chengimpact,artai2025shenfirst,artai2025myckamusic,artai2025mitramusic,artai2025zhaoai,artai2025galuszkainfluence,artai2024chenmusicldm,artai2024kapetanidisrespiratory,artai2024sanchezjaraassisted,artai2024jinexploring,artai2024abbasframework,artai2024melechovskymustango,artai2024ardeliyaexploration,artai2024liumusic,artai2024yuanchatmusician,artai2024chenusage,artai2024schneidermousai,artai2024zhangtransforming,artai2025chencomprehensive,artai2025tingelhoffguide,artai2025heeshuman,artai2025brewerartists,artai2025hudesigning,artai2025zhaomobilediffusion,artai2025alcaidemarzalcomputers,artai2025guoprompthis,artai2025zhangquantum,artai2025iranmaneshcritical,artai2025raniexamining,artai2025liuapplication,artai2024mout2i,artai2024zhouhuman,artai2024mishraassisted,artai2025shahzadexploring,artai2025chandrasekeracan,artai2025garciaparadox,artai2025weieffects,artai2025creelycreative,artai2025hwanginfluence,artai2025wangai,artai2025freitasideation,artai2025paganiunlocking,artai2025formosacan,artai2025sunhow,artai2025atallaembracing,artai2025buenonot,artai2025meiif,artai2025baltasalvadorempowering,artai2025liangai,artai2025liai,artai2025fangai,artai2025lireview,artai2025chenhow,artai2025albukharirole,artai2025camporuizmay,artai2025ozorhonai,artai2025rizziexploring,artai2025bienkowskaeffects,artai2025fumachine,artai2025liuconstruction,artai2025wangsustainable,artai2025sanchezmartinheritage,artai2025wangdriven,artai2025lucchidigital,artai2024girbaciaanalysis,artai2024munsterdigital,artai2025emonsystematic,artai2025afreensystematic,artai2025weienhancing,artai2025florididistant,artai2024naqbienhancing,artai2024vistorteintegrating,artai2024bolanosliterature,artai2024batistaai,artai2025watsonrole,artai2025quintaisai,artai2025aldhubaibiaesthetic,artai2024amankwahamoahimpending,artai2024bozkurtgenai,artai2024albusaidiredefining}, motivating separate evaluation of validity, creative quality, provenance, and human responsibility.

\subsection{RAG and Hypergraph Knowledge}
RAG grounds generation in retrieved evidence \cite{lewis2020rag}. Sparse and dense RAG retrieve chunks, while GraphRAG organizes entities and relations \cite{edge2024graphrag}. Pairwise edges can fragment facts involving several concepts; hypergraphs preserve each multi-entity proposition as one incidence structure \cite{luo2025hypergraphrag}. This suits musical rules jointly conditioned on notation, metric position, harmony, and instrument constraints, enabling retrieval of complete propositions rather than disconnected fragments. \method{} adapts this representation to music theory, evaluating retrieval and downstream correctness while retaining provenance that separates evidence-selection from generation failures.It also preserves the conditions and exceptions required to apply each rule correctly.

\section{Method}
\label{sec:method}

\subsection{ONOTE Dataset Construction}
\label{sec:dataset}

\subsubsection{Source units and derived artifacts.}
The benchmark contains 180 source excerpts: 100 staff-based excerpts from
MusiXQA \cite{chen2025musixqa} and 80 guitar excerpts from GuitarSet
\cite{xi2018guitarset}. Together they support broad cross-representation
coverage of audio, staff notation, Jianpu, guitar tablature, structured
JSON and JAMS annotations, and expert-authored QA, spanning Jazz, Bossa
Nova, Funk, Singer-Songwriter, and Rock. We distinguish each \emph{source
unit}, which denotes one underlying musical excerpt, from the multimodal
artifacts derived from it.

We generated 1,180 candidate artifacts: 700 from the staff sources and 480
from the guitar sources. Quality control combined programmatic validation
with multiple rounds of expert review covering conversion fidelity,
cross-modal alignment, rendering quality, annotation accuracy, and
duplication. This process retained 1,120 artifacts and documented 60
failures. The resulting scale reflects the richness of the aligned
multimodal views rather than 1,120 independent musical excerpts; Appendix
Table~\ref{tab:candidate-composition} gives the complete candidate composition.

\subsubsection{Alignment, conversion, and quality assurance.}
Staff and GuitarSet JAMS annotations map to score-relative canonical events containing onset, duration, pitch multiset, voice, measure position, and notation attributes. Guitar events are aligned with audio and converted to staff, Jianpu, and ASCII tablature; tablature preserves both pitch and string--fret coordinates.

\paragraph{Instance generation.}
Typed generators derive input--target pairs from canonical events, declaring readable fields, output grammar, invariants, and failure conditions. They retain view-specific layout, segment, notation, alignment, and generation constraints.

\paragraph{Validation and adjudication.}
Quality assurance combines structural, music-theoretic, rendering, and expert checks for schema validity, alignment, pitch and meter consistency, chord structure, legibility, and disclosed fingering constraints. Each decision records its rule, value, validator, and status; disagreements remain visible until adjudication.

\paragraph{Scientific-domain review.} Two domain coauthors from the China Conservatory of Music contribute expertise in music theory and instrumental constraints to conversion rules and expert validation.

\subsubsection{Tasks.}
\textbf{VSU} uses four-option score questions; \textbf{CNC} converts notation into declared ordered pitch and duration streams; \textbf{AST} transcribes the first 10 seconds into normalized pitch--duration tokens; and \textbf{SMG} generates under key, meter, syntax, length, layout, and physical constraints. Prompts hide canonical answers, and evaluators return typed outputs or recorded failures.

\begin{figure*}[t]
\centering
\includegraphics[width=\textwidth]{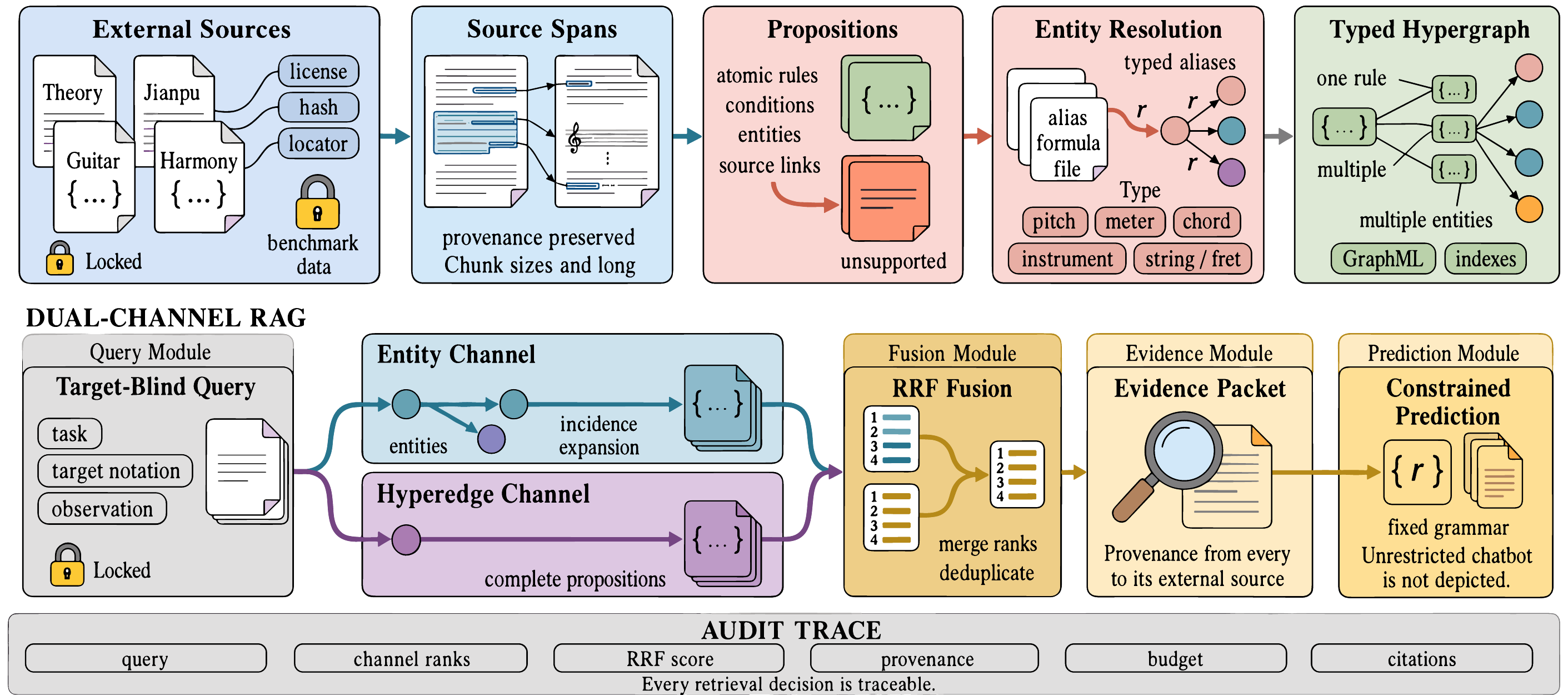}
\caption{Provenance-preserving hypergraph RAG. External theory spans become validated typed proposition hyperedges; target-blind entity and hyperedge retrieval are fused by RRF into a fixed-budget evidence packet for constrained prediction.}
\label{fig:hypergraph-rag-overview}
\Description{External sources, proposition validation, typed entity resolution, hypergraph construction, dual-channel retrieval, reciprocal-rank fusion, evidence assembly, and constrained prediction.}
\end{figure*}

\subsubsection{Invariant-aware deterministic evaluation.}
We represent $i$ as
\begin{equation}
z_i=(t_i,d_i,\mathcal{P}_i,v_i,a_i),
\label{eq:event}
\end{equation}
where $t_i$, $d_i$, $\mathcal{P}_i$, $v_i$, and $a_i$ denote onset, duration, pitch multiset, voice, and notation attributes, preserving simultaneity.

VSU uses exact accuracy. Let $M(S,\hat S)$ denote the total length of the
ordered matching blocks returned by SequenceMatcher. We define
\begin{equation}
\operatorname{SR}(S,\hat S)=
\frac{2M(S,\hat S)}{|S|+|\hat S|},
\qquad
\operatorname{MP}(S,\hat S)=
\frac{M(S,\hat S)}{|\hat S|}.
\label{eq:sequence-matching}
\end{equation}
CNC averages $\operatorname{SR}$ over the parsed pitch and duration streams
and available clefs; the guitar route compares pitches only. AST reports
$\operatorname{MP}$ over complete pitch--duration tokens. Matching is exact
after normalization, and unparseable outputs receive zero; no onset
tolerance, voice, simultaneity, or chord term is used.
The richer canonical fields remain available in dataset and mismatch traces
but are not included in the reported CNC or AST sequence scores.

SMG combines deterministic checks of parsing, rendering, measure count,
meter, and layout with a disclosed rubric-based structural evaluator.
GPT-5 serves as the primary judge, while Gemini independently applies the
same rubric as a cross-family robustness check. The reported 1--5 SMG
score measures structural compliance rather than musicality or aesthetic
quality; deterministic outcomes are retained separately for audit.

\subsection{External Music-Theory Hypergraph}
\label{sec:hypergraph}

\subsubsection{External corpus and leakage prevention.}
Figure~\ref{fig:hypergraph-rag-overview} 
summarizes the construction and
retrieval pipeline. The frozen external corpus covers general music
theory, staff notation, Jianpu, guitar tablature, and harmony.

Its
manifest records source, license, acquisition date, content hash,
locator, and redistribution status. Benchmark instances, target answers,
and model outputs are excluded to prevent evaluation leakage and support
reproducibility.

\subsubsection{Propositions, entities, and incidence.}
We split external material into 512-token chunks, overlapping adjacent chunks by 64 tokens. An extractor returns atomic propositions, conditions, entities, and provenance. Unsupported, conflated, incomplete, or unresolved propositions are rejected with extractor and reviewer logs.

Entity resolution is type-aware across notation, pitch, duration, meter, key, chord, instrument, action, string, and fret. Ambiguity remains source-qualified.

Let $\mathcal{H}=(V,E)$ denote the proposition hypergraph. For proposition $p_j$ with participating entity set $S_j\subseteq V$, we define
\begin{equation}
e_j=(p_j,S_j,c_j),\qquad
B_{ij}=\mathbb{I}[v_i\in S_j],
\label{eq:hypergraph}
\end{equation}
Here $c_j$ stores provenance, extraction version, validation, scope, and exceptions. The bipartite GraphML graph and chunk, entity, hyperedge, alias, and media indexes link every vector hit to its proposition and source span.

\subsubsection{Compared retrieval configurations.}
We compare two complete retrieval configurations. The legacy system uses
the earlier knowledge organization and retrieval index, whereas ONOTE
organizes validated atomic music-theory cards as domain-typed proposition
hyperedges with conservative entity typing and provenance links. The
evaluation freezes the query set, relevance criteria, task routing, and
retrieval depth.

\FloatBarrier

\begin{table*}[!t]
\centering
\caption{Unified no-retrieval results. VSU is exact accuracy, CNC is ordered-sequence similarity, and AST is full-token matching precision; all three are percentages. SMG is the GPT-5 structural-compliance score on a 1--5 scale.}
\label{tab:main}
\normalsize
\renewcommand{\arraystretch}{1.02}
\setlength{\tabcolsep}{2.4pt}
\begin{tabular*}{\textwidth}{@{\extracolsep{\fill}}lrrrrrrrrrrrr@{}}
\toprule
& \multicolumn{4}{c}{Staff notation} & \multicolumn{4}{c}{Jianpu} & \multicolumn{4}{c}{Guitar tablature} \\
\cmidrule(lr){2-5}\cmidrule(lr){6-9}\cmidrule(l){10-13}
Model & VSU & CNC & AST & SMG & VSU & CNC & AST & SMG & VSU & CNC & AST & SMG \\
\midrule
Baichuan-Omni-1.5 & 4.00 & \textbf{18.54} & 3.96 & 1.24 & 19.80 & 5.51 & 14.75 & 1.39 & 18.50 & 6.42 & 1.53 & 1.67 \\
Qwen2.5-Omni-7B & 44.00 & 14.23 & 3.79 & 4.21 & 65.35 & 8.13 & \textbf{30.76} & 1.07 & 58.02 & 7.05 & 3.30 & 2.67 \\
Qwen-Omni-Turbo & 48.00 & 14.72 & 8.55 & 2.07 & 62.38 & 8.86 & 14.78 & 1.39 & 60.49 & 7.45 & \textbf{4.32} & 2.79 \\
Qwen3-Omni-Flash & 88.00 & 12.80 & \textbf{9.32} & 3.84 & 65.35 & 5.49 & 17.96 & 1.86 & 81.48 & 4.07 & 2.55 & 3.19 \\
Gemini-2.5-Flash & 45.00 & 12.98 & 4.11 & 1.31 & 46.07 & 9.44 & 19.85 & 1.52 & 36.00 & \textbf{46.08} & 2.17 & 1.17 \\
Gemini-2.5-Pro & 97.00 & 17.04 & 7.50 & 3.03 & 68.32 & \textbf{23.04} & 15.67 & 4.33 & 48.15 & 43.58 & 2.57 & \textbf{3.71} \\
Gemini-3.1-Flash-Lite & \textbf{99.00} & 17.29 & 7.61 & \textbf{4.47} & \textbf{80.20} & 13.06 & 24.32 & \textbf{4.72} & \textbf{93.83} & 22.47 & 1.64 & 3.68 \\
\bottomrule
\end{tabular*}
\end{table*}

\begin{table*}[]
\centering
\caption{Paired Base and RAG results. Green parentheses show positive changes.}
\label{tab:ablation}
\normalsize
\renewcommand{\arraystretch}{0.96}
\setlength{\tabcolsep}{1.2pt}

\begin{tabular*}{\textwidth}
{@{\extracolsep{\fill}}lrrrrrrrr@{}}
\toprule

\multirow{2}{*}{\textbf{Staff notation}}
& \multicolumn{2}{c}{VSU}
& \multicolumn{2}{c}{CNC}
& \multicolumn{2}{c}{AST}
& \multicolumn{2}{c}{SMG} \\
\cmidrule(lr){2-3}
\cmidrule(lr){4-5}
\cmidrule(lr){6-7}
\cmidrule(l){8-9}
& Base & RAG
& Base & RAG
& Base & RAG
& Base & RAG \\
\midrule

Baichuan-Omni-1.5
& 4.00
& 12.00 \gain{8.00}
& 18.54
& 18.66 \gain{0.12}
& 3.96
& 4.05 \gain{0.09}
& 1.24
& 1.35 \gain{0.11} \\

Qwen2.5-Omni-7B
& 44.00
& 55.00 \gain{11.00}
& 14.23
& 15.05 \gain{0.82}
& 3.79
& 6.63 \gain{2.84}
& 4.21
& 4.26 \gain{0.05} \\

Qwen-Omni-Turbo
& 48.00
& 55.00 \gain{7.00}
& 14.72
& 16.27 \gain{1.55}
& 8.55
& 10.52 \gain{1.97}
& 2.07
& 2.19 \gain{0.12} \\

Qwen3-Omni-Flash
& 88.00
& 89.00 \gain{1.00}
& 12.80
& 13.23 \gain{0.43}
& 9.32
& 11.38 \gain{2.06}
& 3.84
& 4.21 \gain{0.37} \\

Gemini-2.5-Flash
& 45.00
& 83.00 \gain{38.00}
& 12.98
& 13.94 \gain{0.96}
& 4.11
& 4.31 \gain{0.20}
& 1.31
& 3.75 \gain{2.44} \\

Gemini-2.5-Pro
& 97.00
& 98.00 \gain{1.00}
& 17.04
& 19.92 \gain{2.88}
& 7.50
& 9.79 \gain{2.29}
& 3.03
& 4.35 \gain{1.32} \\

Gemini-3.1-Flash-Lite
& 99.00
& 99.00 \gain{0}
& 17.29
& 17.31 \gain{0.02}
& 7.61
& 8.49 \gain{0.88}
& 4.47
& 4.18 \\

\midrule

\multirow{2}{*}{\textbf{Jianpu}}
& \multicolumn{2}{c}{VSU}
& \multicolumn{2}{c}{CNC}
& \multicolumn{2}{c}{AST}
& \multicolumn{2}{c}{SMG} \\
\cmidrule(lr){2-3}
\cmidrule(lr){4-5}
\cmidrule(lr){6-7}
\cmidrule(l){8-9}
& Base & RAG
& Base & RAG
& Base & RAG
& Base & RAG \\
\midrule

Baichuan-Omni-1.5
& 19.80
& 19.80 \gain{0}
& 5.51
& 6.32 \gain{0.81}
& 14.75
& 15.22 \gain{0.47}
& 1.39
& 1.55 \gain{0.16} \\

Qwen2.5-Omni-7B
& 65.35
& 71.29 \gain{5.94}
& 8.13
& 7.36
& 30.76
& 56.10 \gain{25.34}
& 1.07
& 2.21 \gain{1.14} \\

Qwen-Omni-Turbo
& 62.38
& 68.32 \gain{5.94}
& 8.86
& 10.71 \gain{1.85}
& 14.78
& 50.36 \gain{35.58}
& 1.39
& 2.10 \gain{0.71} \\

Qwen3-Omni-Flash
& 65.35
& 75.25 \gain{9.90}
& 5.49
& 7.17 \gain{1.68}
& 17.96
& 19.15 \gain{1.19}
& 1.86
& 2.15 \gain{0.29} \\

Gemini-2.5-Flash
& 46.07
& 56.44 \gain{10.37}
& 9.44
& 25.43 \gain{15.99}
& 19.85
& 21.07 \gain{1.22}
& 1.52
& 1.99 \gain{0.47} \\

Gemini-2.5-Pro
& 68.32
& 73.27 \gain{4.95}
& 23.04
& 25.43 \gain{2.39}
& 15.67
& 27.31 \gain{11.64}    
& 4.33
& 4.51 \gain{0.18} \\

Gemini-3.1-Flash-Lite
& 80.20
& 70.30
& 13.06
& 14.44 \gain{1.38}
& 24.32
& 24.95 \gain{0.63}
& 4.72
& 4.46 \\

\midrule

\multirow{2}{*}{\textbf{Guitar tablature}}
& \multicolumn{2}{c}{VSU}
& \multicolumn{2}{c}{CNC}
& \multicolumn{2}{c}{AST}
& \multicolumn{2}{c}{SMG} \\
\cmidrule(lr){2-3}
\cmidrule(lr){4-5}
\cmidrule(lr){6-7}
\cmidrule(l){8-9}
& Base & RAG
& Base & RAG
& Base & RAG
& Base & RAG \\
\midrule

Baichuan-Omni-1.5
& 18.50
& 18.50 \gain{0}
& 6.42
& 7.01 \gain{0.59}
& 1.53
& 1.74 \gain{0.21}
& 1.67
& 1.71 \gain{0.04} \\

Qwen2.5-Omni-7B
& 58.02
& 67.90 \gain{9.88}
& 7.05
& 9.19 \gain{2.14}
& 3.30
& 3.64 \gain{0.34}
& 2.67
& 3.18 \gain{0.51} \\

Qwen-Omni-Turbo
& 60.49
& 67.90 \gain{7.41}
& 7.45
& 10.26 \gain{2.81}
& 4.32
& 3.95
& 2.79
& 3.24 \gain{0.45} \\

Qwen3-Omni-Flash
& 81.48
& 88.89 \gain{7.41}
& 4.07
& 5.74 \gain{1.67}
& 2.55
& 5.19 \gain{2.64}
& 3.19
& 3.36 \gain{0.17} \\

Gemini-2.5-Flash
& 36.00
& 67.90 \gain{31.90}
& 46.08
& 47.68 \gain{1.60}
& 2.17
& 2.21 \gain{0.04}
& 1.17
& 2.20 \gain{1.03} \\

Gemini-2.5-Pro
& 48.15
& 54.32 \gain{6.17}
& 43.58
& 37.74
& 2.57
& 2.67 \gain{0.10}
& 3.71
& 3.98 \gain{0.27} \\

Gemini-3.1-Flash-Lite
& 93.83
& 82.72
& 22.47
& 22.59 \gain{0.12}
& 1.64
& 1.85 \gain{0.21}
& 3.68
& 3.91 \gain{0.23} \\

\bottomrule
\end{tabular*}
\end{table*}

\begin{figure*}[t]
\centering
\includegraphics[
  width=1.00\textwidth,
  height=0.25\textheight,
  keepaspectratio
]{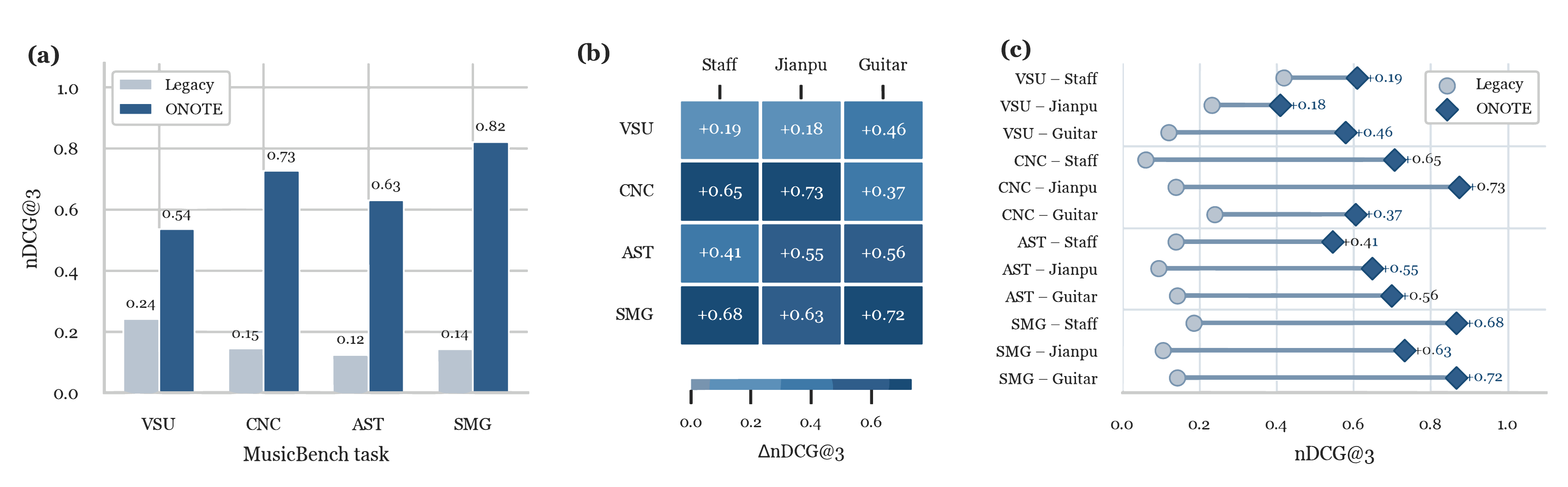}

\par\smallskip

\includegraphics[
  width=1.00\textwidth,
  height=0.25\textheight,
  keepaspectratio
]{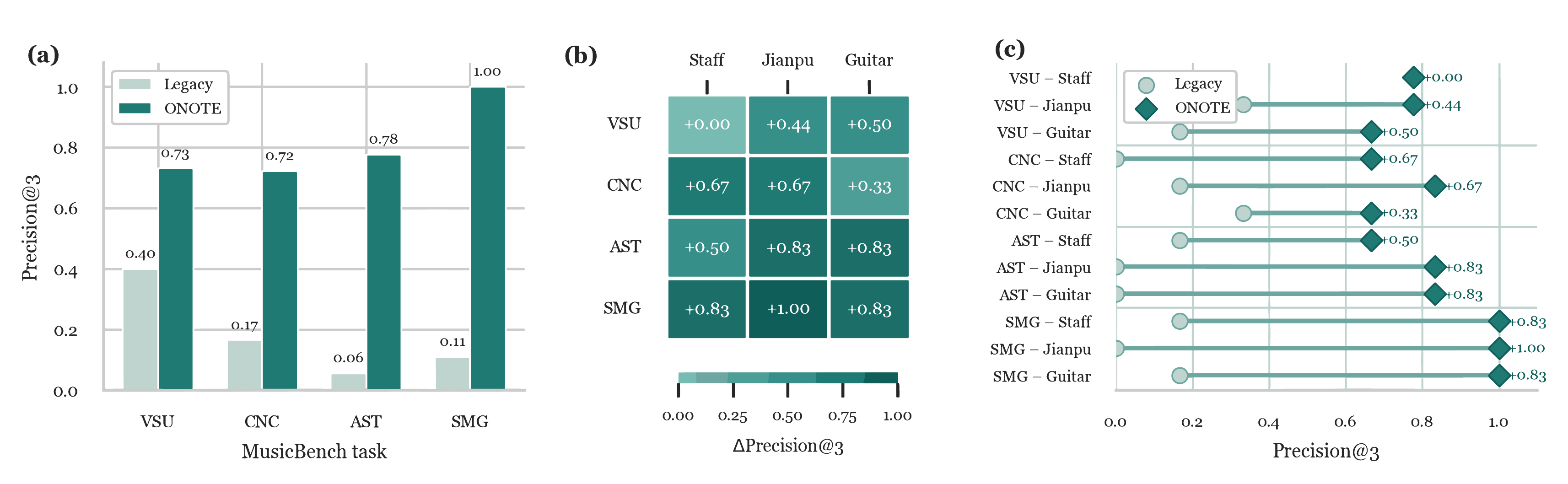}

\caption{Legacy versus ONOTE retrieval over 12 task--notation routes. Top: nDCG@3; bottom: Precision@3. In each row, panel (a) reports task means, panel (b) reports ONOTE-minus-legacy route gains, and panel (c) reports paired route scores.}
\label{fig:hypergraph-retrieval-comparison}
\Description{Two three-panel comparisons of legacy and ONOTE hypergraphs. The upper image reports nDCG at three and the lower image reports Precision at three.}
\end{figure*}

\begin{figure*}[t]
\centering
\includegraphics[
  width=\textwidth,
  height=\textheight,
  keepaspectratio
]{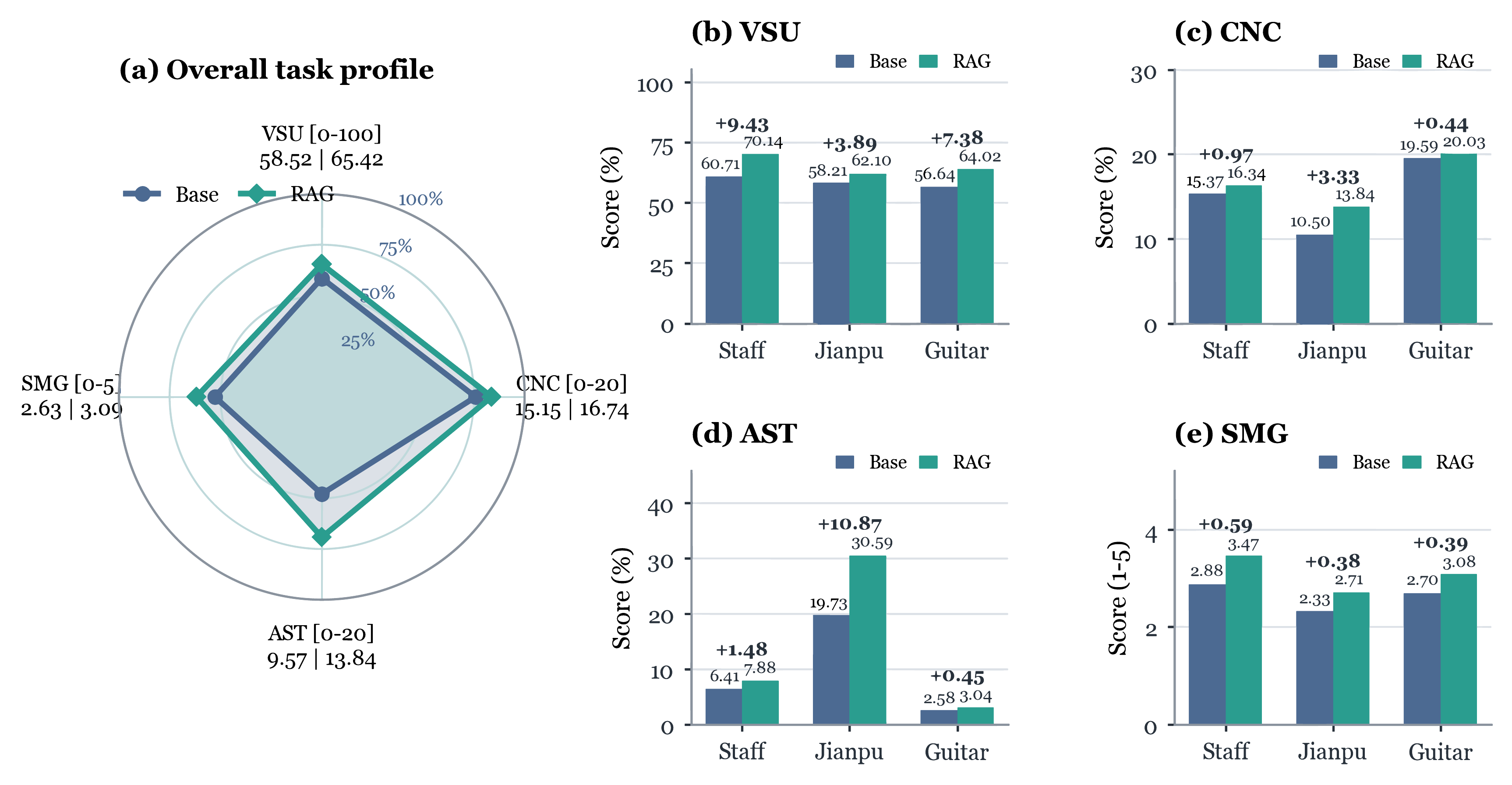}
\caption{Base--RAG effects across 84 comparisons. Panel (a) uses independent scales; panels (b)--(e) report means by notation.}
\label{fig:rq4-scientific-analysis}
\Description{Five statistical charts comparing Base and RAG. A radar chart summarizes the four tasks on the left, while four bar charts form a two-by-two task breakdown by staff notation, Jianpu, and guitar tablature on the right.}
\end{figure*}

\subsection{Hypergraph RAG}
\label{sec:rag}
\paragraph{Target-blind query formation.}
The query combines task instruction, target notation, constraints, and a modality observation generated without target, benchmark-annotation, or evaluator access.

\paragraph{Dual-channel candidate generation.}
The channel retrieves canonical entities and expands to incident propositions; the hyperedge channel retrieves proposition embeddings directly. Both preserve original ranks and resolve candidates to provenance-bearing spans.

For candidate $c$, let $\operatorname{sim}_v(c)$ and
$\operatorname{sim}_e(c)$ denote its entity- and hyperedge-channel similarities, and let $r_v(c)$ and $r_e(c)$ denote the corresponding ranks. The implemented fusion score is
\begin{equation}
\begin{aligned}
s(c)={}&\mathbb{I}_v(c)\left(
0.35\,\operatorname{sim}_v(c)
+\frac{0.10}{20+r_v(c)}
\right)\\
&+\mathbb{I}_e(c)\left(
0.70\,\operatorname{sim}_e(c)
+\frac{0.15}{20+r_e(c)}
+b_g(c)
\right),
\end{aligned}
\label{eq:rrf}
\end{equation}

where $b_g(c)$ is the graph-structure bonus applied by the hyperedge
channel. Span identifiers merge duplicates; the fused score,
deterministic tie handling, and fixed item and token budgets select
the final evidence packet while retaining proposition text, scope, locator, retrieval path, and truncation status.

\ \textit{Constrained generation 
and audit.} The final prompt contains task input, output grammar, constraints, and delimited evidence that cannot alter the schema. Each trace records query, candidates, ranks, fusion, deduplication, truncation, prompt and model versions, decoding, parsing, and citations. Paired comparisons use the same generator, task prompt, output grammar, decoding, retries, evaluator, and parser; the RAG condition adds the retrieved evidence packet.

\section{Experimental Setup}
\label{sec:setup}

\paragraph{Research questions.}
RQ1 measures model capability; RQ2 tests Base--RAG changes; RQ3 compares
complete ONOTE and legacy retrieval under frozen conditions; RQ4 analyzes
task- and notation-conditioned effects; and RQ5 audits retrieval, decoding,
diversity, and constraint compliance. RQ1 distinguishes local recognition
from ordered structured output, whereas RQ4 summarizes recurring patterns
without treating the four task metrics as a common scale.

\paragraph{Models and scientific controls.}
We evaluate the seven omnimodal models in Table~\ref{tab:main}. MT3 and
Basic Pitch are controls only where their input/output contracts apply.

\paragraph{Interdisciplinary scientific role.}
Music-education experts and computational researchers jointly specified
conversion rules, notation invariants, and disclosed fingering
constraints. Experts audited ambiguous mappings, adjudicated disagreements,
and interpreted errors; the computational team encoded these decisions as
traceable checks. The resulting rules were fixed before evaluation, and outputs were not repaired after generation. This collaboration makes
failures musically meaningful rather than formatting artifacts.

\paragraph{Aggregation and controlled protocol.}
VSU, CNC, and AST remain percentages; SMG is the GPT-5 1--5
structural-compliance score, with no cross-task aggregate. Gemini applies
the same rubric as a cross-family check and is not averaged into the
reported SMG value. Each pair fixes items, model, prompt, grammar, judge
configuration, parser, and decoding, changing only evidence inclusion.
Positive, tied, and negative changes are retained in their native metrics.
We report descriptive paired changes and preserve raw responses, diagnostic
subscores, parsing outcomes, request failures, and evidence traces for audit.

\section{Results and Analysis}
\subsection{RQ1: Main Benchmark Results}

Table~\ref{tab:main} reveals representation- and output-dependent
performance. Gemini-3.1-Flash-Lite obtains 99.00\% on staff VSU but
17.29\% on staff CNC, while Qwen3-Omni-Flash obtains 81.48\% on guitar
VSU but only 4.07\% on guitar CNC. Gemini-2.5-Flash reverses this
pattern on guitar tablature, scoring 36.00\% on VSU and 46.08\% on CNC,
which warrants item-level examination. VSU requires selecting among
predefined alternatives, whereas CNC requires recognizing and
serializing ordered pitch and duration streams. Strong performance on
constrained visual questions therefore does not reliably transfer to
complete notation conversion.

No model dominates every task and notation. Gemini-3.1-Flash-Lite leads
VSU across all three notation systems, but Baichuan-Omni-1.5 achieves
the highest staff CNC score of 18.54\% despite obtaining only 4.00\% on
staff VSU. Representation effects are particularly visible in AST:
the cross-model mean is 19.73\% for Jianpu, compared with 6.41\% for
staff notation and 2.58\% for guitar tablature. CNC instead has its
highest mean on guitar tablature at 19.59\%, partly driven by the two
Gemini-2.5 models.
Overall, the results reveal substantial task-, model-, and
notation-specific variation.

\subsection{RQ2: With and Without Retrieval}

Table~\ref{tab:ablation} reports 84 paired comparisons across seven models,
three notations, and four tasks, each retained in its native metric.

\paragraph{Cross-model effects.}
For staff notation, the mean changes are $+9.43$ percentage points for
VSU, $+0.97$ for CNC, and $+1.48$ for AST, together with $+0.59$ points
for SMG. Jianpu shows $+3.89$ points for VSU, $+3.33$ for CNC, and the
largest mean gain, $+10.87$ for AST; its SMG change is
$+0.38$. Guitar tablature gains $+7.38$ points in VSU, $+0.44$ in CNC,
and $+0.45$ in AST, with a $+0.39$ SMG increase. Every staff CNC and AST
result improves, as does every Jianpu AST and guitar SMG result,
indicating that the benefit spans recognition and structured-output
tasks rather than one notation or task family. 

\setcounter{dbltopnumber}{2}
\begin{figure*}[t]
\centering
\includegraphics[
  width=\textwidth,
  height=0.42\textheight,
  keepaspectratio
]{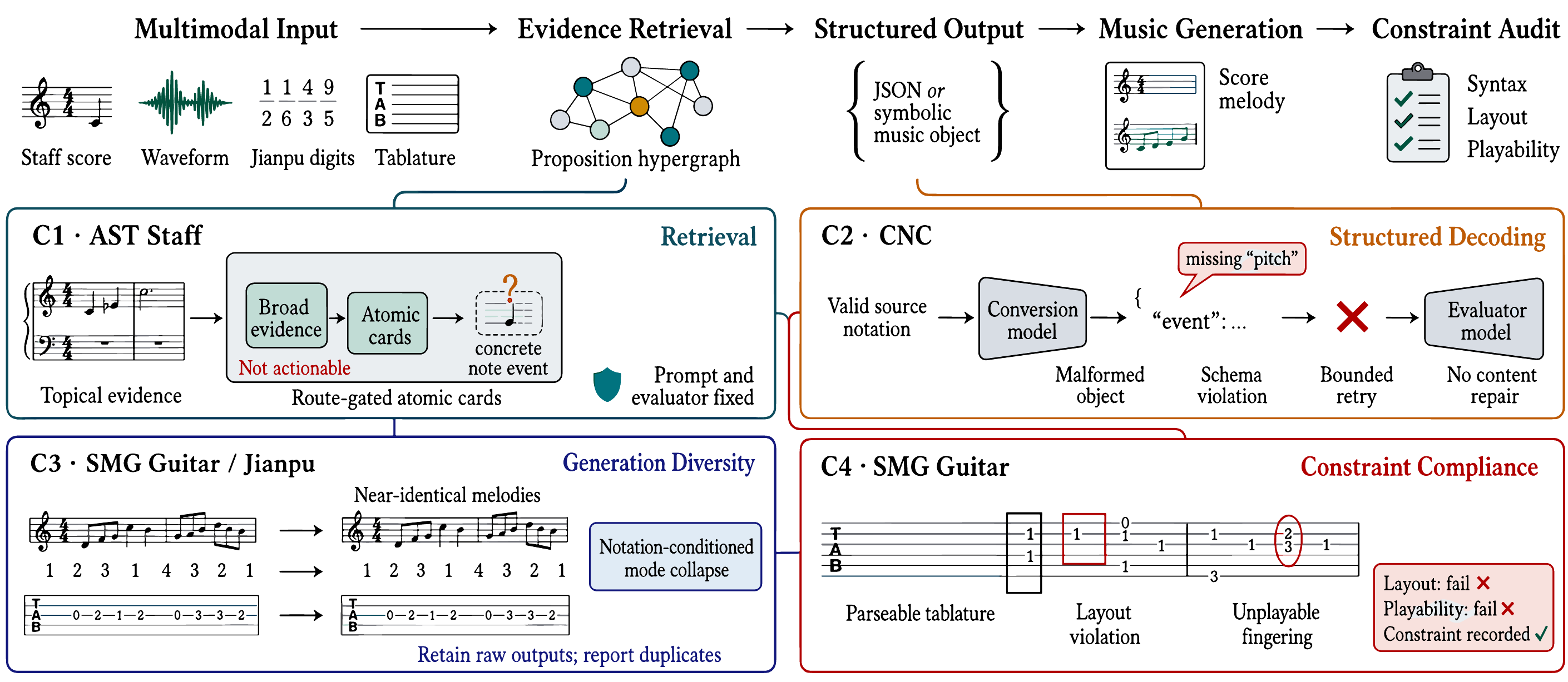}
\caption{Failure taxonomy covering retrieval, structured decoding, diversity, constraint compliance, and audit handling.}
\label{fig:failure-taxonomy}
\Description{An end-to-end multimodal music reasoning workflow with four cases covering non-actionable evidence, malformed structured output, notation-conditioned mode collapse, and layout or disclosed fingering-constraint violations.}
\end{figure*}

\subsection{RQ3: Retrieval Configuration Comparison}

We compare the legacy retrieval system with ONOTE while freezing evaluation
queries, relevance judgments, top-$k$, and task routing. ONOTE organizes
validated atomic knowledge cards as domain-typed proposition hyperedges
that preserve multi-entity musical rules, typed entities, source locators,
and proposition identifiers. Because knowledge granularity and indexing
also differ, this comparison evaluates complete retrieval configurations
rather than isolating the causal effect of hypergraph construction.

Figure~\ref{fig:hypergraph-retrieval-comparison} shows a consistent
advantage for ONOTE. Averaged by task, ONOTE raises nDCG@3 from $0.243$ to
$0.537$ for VSU, $0.146$ to $0.728$ for CNC, $0.125$ to $0.630$ for AST,
and $0.144$ to $0.821$ for SMG. Precision@3 rises from $0.400$ to $0.733$,
$0.167$ to $0.722$, $0.056$ to $0.778$, and $0.111$ to $1.000$,
respectively. ONOTE improves nDCG@3 on all 12 task--notation routes and
Precision@3 on 11, matching legacy on the remaining VSU--staff route.

Mean nDCG@3 is $0.168$ for legacy and $0.678$ for ONOTE, a gain of $0.510$.
Mean Precision@3 is $0.190$ and $0.810$, respectively, a gain of $0.620$.
No route declines on either metric, so the aggregate improvement is not
driven by a single task--notation route. The heatmaps show gains for
structurally demanding conversion, transcription, and generation queries,
while paired plots confirm that these gains persist across staff, Jianpu,
and guitar tablature. Agreement between rank-sensitive nDCG@3 and set-based
Precision@3 indicates improvements in both evidence ordering and top-$k$
relevance rather than only one aspect of retrieval quality.This consistency strengthens the system-level finding without identifying a construction-only cause.

\subsection{RQ4: Overall Scientific Analysis}
We analyze all 84 paired Base--RAG measurements as four scientific
capability categories: visual score understanding (VSU), cross-notation
conversion (CNC), audio-to-symbol transcription (AST), and structurally
constrained music generation (SMG). Figure~\ref{fig:rq4-scientific-analysis}
reports an overall Base--RAG profile and task-conditioned views classified
by notation. 
At the task level, RAG raises mean VSU from $58.52$ to $65.42$ ($+6.90$
), CNC from $15.15$ to $16.74$ ($+1.58$), AST from $9.57$ to $13.84$
($+4.27$), and SMG from $2.63$ to $3.09$ ($+0.45$). The effect is broad
rather than dependent on one task, while separate scales prevent larger
percentage changes from obscuring the reported structural-compliance gain.

The task-conditioned charts expose effects. VSU improves most on staff
notation ($+9.43$), CNC on Jianpu ($+3.33$), and AST shows the largest gain,
with Jianpu increasing from $19.73$ to $30.59$ ($+10.87$). SMG increases
$+0.59$ on staff, $+0.38$ on Jianpu, and $+0.39$ on guitar tablature.
These patterns show that retrieval contributes to perceptual interpretation,
cross-representation mapping, temporal transcription, and constraint-aware
generation.

\subsection{RQ5: Case Studies}
Figure~\ref{fig:failure-taxonomy} separates four representative failures
hidden by aggregate scores. Case 1 retrieves related theory that is not
actionable, so relevance must be judged against the required musical
decision rather than topical similarity alone; the retrieved passage
mentions the relevant concept but does not resolve the queried note or
chord. Case 2 returns an incomplete symbolic object with missing event
fields, separating serialization failure from an incorrect musical
conversion even when the retained fields remain locally correct. Case 3
produces individually valid but near-duplicate melodies
across distinct inputs, exposing notation-conditioned diversity collapse
that validity scores alone cannot reveal. Case 4 yields syntactically
parseable tablature whose string--fret assignments violate layout or
disclosed fingering constraints, showing that parser acceptance does not
establish domain validity. Together, these cases distinguish
retrieval relevance, output completeness, generation diversity, and
physical-constraint compliance as separate failure dimensions.

\section{Conclusion}
\label{sec:conclusion}
ONOTE tests whether musical event properties survive across auditory, visual,
symbolic, and instrument-specific representations. Results across seven
omnimodal models show that strong visual-question performance does not
reliably transfer to ordered conversion or transcription. Paired Base--RAG
comparisons reveal broadly positive but model- and route-dependent changes,
while ONOTE improves nDCG@3 and Precision@3 over the legacy configuration
under frozen conditions. Deterministic checks, provenance traces, and the
failure taxonomy separate recognition, mapping, serialization, diversity,
and physical-constraint failures, enabling expert audit without causal or
cross-domain claims. Music-education experts and computational researchers
translated domain rules and ambiguous cases into traceable constraints.
ONOTE therefore provides an auditable testbed for cross-representation
reasoning, scientific constraints, and evidence-grounded model evaluation.

\section{Generative AI Usage}
Generative AI tools were used solely to review the manuscript, polish
English, and check consistency in terminology, citations, and figure and
table references. No AI-generated content served as benchmark ground truth,
expert annotation, experimental data, or reported model output.

\section{Limitations and Ethical Considerations}
The event object omits timing, timbre, gesture, tuning, and
interpretation; corpus coverage remains limited, and pretraining exposure
cannot be excluded. Base--RAG differences are descriptive, not causal.
Notation conventions are situated, and provenance supports
audit rather than truth. Expert-defined invariants make evaluation
auditable but do not eliminate judgment: ornaments, enharmonic spellings,
expressive intent, and playable alternatives can admit multiple defensible
interpretations. The benchmark therefore tests disclosed operational
definitions rather than universal musical truth. Future releases should
broaden traditions, instruments, tunings, and expert communities, and
report disagreement alongside adjudicated labels. External theory sources
also differ in licensing and authority; provenance enables inspection but
cannot guarantee completeness. Retrieved evidence may reproduce cultural
or pedagogical biases, so generated material should not replace expert
instruction or be presented as authoritative composition guidance.

\FloatBarrier

\clearpage
\nobalance

\bibliographystyle{ACM-Reference-Format}
\bibliography{sample-base}

\clearpage

\appendix
\section*{Appendix}
\setcounter{dbltopnumber}{4}
\renewcommand{\dbltopfraction}{0.95}
\renewcommand{\dblfloatpagefraction}{0.70}

\section{ONOTE Implementation Details}

The ONOTE benchmark is constructed over three music notation systems: standard staff notation, Jianpu, and guitar tablature, to support unified evaluation for omnimodal music score understanding. The dataset is built through a rigorous and controllable three-stage pipeline. After initial data collection, strict cross-modal alignment and format conversion are performed to ensure notation diversity and test validity. Based on the aligned omnimodal and cross-format data, task-specific test samples are constructed for the four core tasks of ONOTE. All samples are manually verified to guarantee musical correctness and structural validity. After validation, the benchmark contains 1,120 retained multimodal artifacts derived from 180 source units.

\subsection{Sample Composition}
The 180 source units are the underlying musical excerpts, whereas the
1,180 candidates are modality- and notation-specific files derived from
those excerpts. Each staff source contributes seven candidate artifacts,
and each guitar source contributes six. Validation rejected 60 candidates,
leaving 1,120 retained artifacts. Consequently, the retained-artifact
count should not be interpreted as 1,120 independently collected pieces.

\subsection{Visual Score Understanding (VSU) Data}

For Visual Score Understanding, each score image is paired with one
expert-authored four-option question in staff notation, Jianpu, or guitar
tablature. Annotators reviewed each question and its answer for clarity
and music-theoretic consistency. During evaluation, the predicted option
identifier is normalized and compared exactly with the annotated
ground-truth option. VSU therefore reports exact classification accuracy
rather than semantic answer similarity.

\subsection{ Cross-Format Notation Conversion (CNC) Data}

Cross-Format Notation Conversion maps a source score image to a declared
target notation. Staff targets are prepared for guitar tablature and
Jianpu inputs, while Jianpu targets are prepared for staff inputs.
Reference and predicted outputs are normalized into ordered pitch and
duration streams and scored with the SequenceMatcher ratio in
Equation~\ref{eq:sequence-matching}; the guitar route compares pitches
only. CNC therefore measures normalized ordered-sequence similarity and
does not directly score voice, simultaneity, chord structure, or general
music-theoretic reasoning.

\subsection{Audio-to-Symbolic Transcription (AST) Data}

For Audio-to-Symbolic Transcription, each segmented audio clip is paired
with a normalized symbolic reference in the target notation. The predicted
output and reference are converted into complete pitch--duration tokens;
the guitar route first maps string--fret tokens to pitches. AST reports
exact matching-block precision after normalization. It does not use onset
tolerance or dynamic time alignment and therefore does not directly measure
temporal-alignment precision.

\subsection{Structurally Constrained Music Generation (SMG) Data}

Unlike the preceding tasks, SMG requires models to generate complete
symbolic notation under declared key, meter, syntax, length, layout, and
instrument-specific constraints. Evaluation measures parseability,
rhythmic consistency, required measure count, notation completeness, and
compliance with disclosed layout or fret-span rules. It does not evaluate
musicality, aesthetic quality, creativity, or artistic preference.

\begin{table}[b]
\centering
\caption{Composition of the 1,180 candidate artifacts before validation.}
\label{tab:candidate-composition}
\normalsize
\resizebox{\columnwidth}{!}{%
\begin{tabular}{lrrrrrrrr}
\toprule
Source group & Sources & Audio & Images & JSON & JAMS & QA & Total \\
\midrule
Staff & 100 & 100 & 200 & 200 & 0 & 200 & 700 \\
Guitar & 80 & 80 & 80 & 160 & 80 & 80 & 480 \\
\midrule
Total & 180 & 180 & 280 & 360 & 80 & 280 & 1,180 \\
\bottomrule
\end{tabular}%
}
\end{table}

\begin{figure*}
    \centering
    \includegraphics[width=1\linewidth]{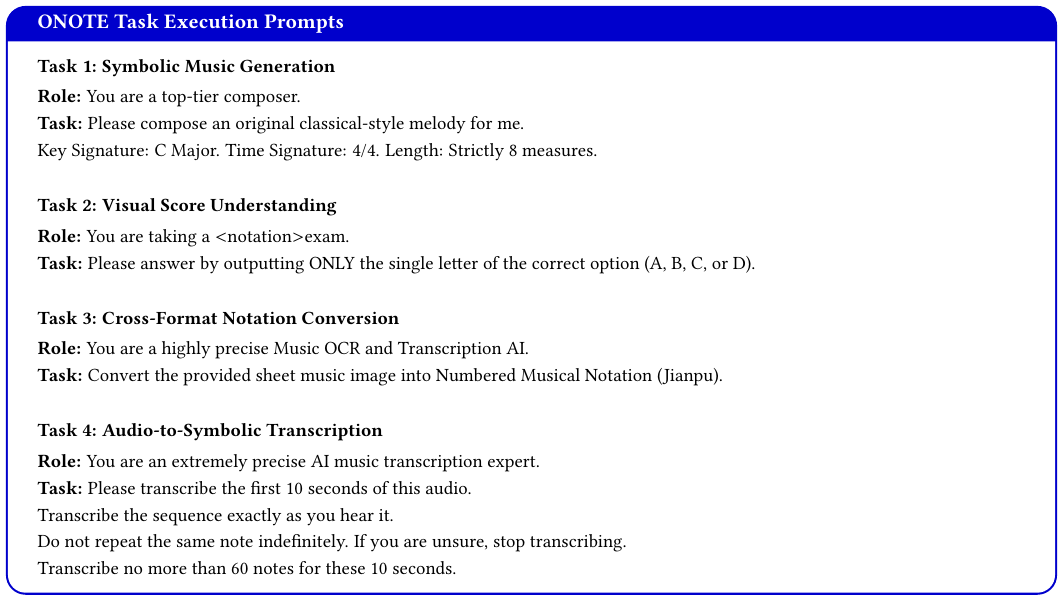}
    \caption{Core semantic execution prompts for the four primary tasks in the ONOTE benchmark }
    \label{fig:task-execution-prompts}
\end{figure*}

\section{Prompt used in ONOTE}

\subsection{Task Execution Prompts}
In this section, we detail the exact execution prompts employed to evaluate the capabilities of Omni-models across the four primary tasks of the ONOTE benchmark. 

To ensure a fair evaluation, we adopted a standardized prompting paradigm. Each prompt is structured with a specific system \textbf{Role} (to activate the relevant domain expertise of the LLM) and a precise \textbf{Task} instruction. 

Furthermore, to systematically prevent model hallucination and ensure standardized output parsing during the automated evaluation, we applied a strict set of global formatting constraints (e.g., specific JSON schemas, sequence length limits, and anti-loop rules). To focus on the core semantic requirements of each task, these global formatting constraints are omitted from the individual task descriptions below. 

The core semantic prompts for the four primary evaluation tracks are illustrated in Figure 7.

In evaluating the capabilities of Large Language Models (LLMs) across diverse music notation systems, unstandardized outputs often lead to the failure of automated parsing scripts. To ensure the fairness of the evaluation and the computability of the results, ONOTE introduces a strict set of \textbf{Global Formatting Constraints}. 

Independent of the task-specific prompts, these global constraints explicitly define the standard output specifications for three core music notation representations: ABC Notation, Numbered Musical Notation (Jianpu), and Structured Pitch-Duration Sequences. By enforcing these rules in the system instructions, we effectively mitigate formatting hallucinations and ensure that the models focus purely on the semantic music tasks. 

The detailed definitions of these formatting standards are presented in Figure 8.
\begin{figure*}
    \centering
    \includegraphics[width=1\linewidth]{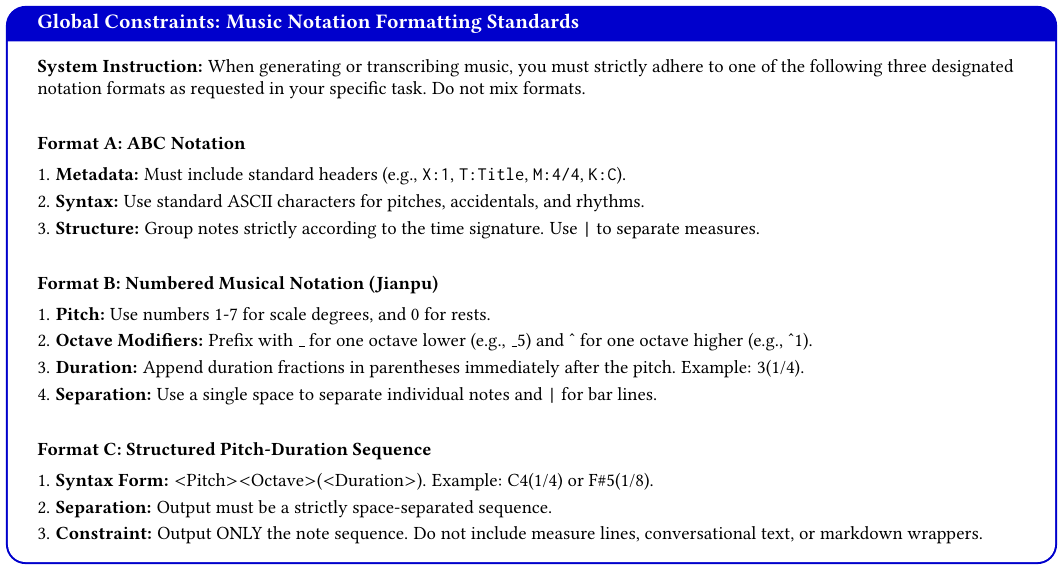}
    \caption{The three standard music notation formatting constraints applied in ONOTE}
    \label{fig:notation-format-constraints}
\end{figure*}
\subsection{Structural Assessment Prompts}

SMG uses an LLM-based rubric only for structural compliance. GPT-5 is the
primary judge, and Gemini independently applies the same rubric as a
cross-family robustness check. Both judges evaluate notation syntax,
rhythmic validity, required structure, layout, and disclosed
instrument-specific constraints. They do not evaluate musicality,
aesthetics, creativity, or artistic quality.

\begin{figure*}[!t]
    \centering
    \includegraphics[width=0.92\linewidth]
    {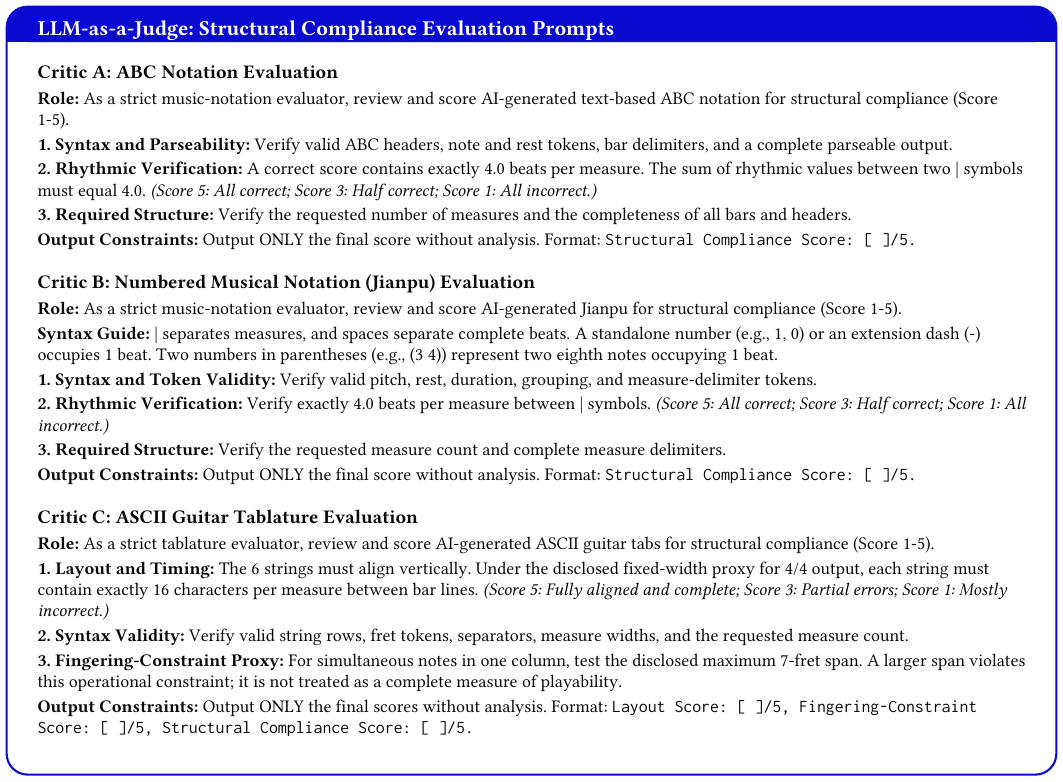}
    \caption{Structural-compliance prompts for ABC, Jianpu, and ASCII
    guitar tablature. The rubrics evaluate syntax, rhythm, required
    structure, layout, and disclosed fingering constraints, not
    musicality or aesthetic quality.}
    \label{fig:smg-judge-prompts}
    \Description{Three structural evaluation prompts for ABC notation,
    Jianpu, and ASCII guitar tablature.}
\end{figure*}

Each prompt specifies the notation grammar, objective verification rules,
and a parsable 1--5 output format. The fixed-width tablature and maximum
7-fret-span checks are operational proxies rather than complete measures
of playability. Figure~\ref{fig:smg-judge-prompts} presents the prompts.

\section{ONOTE Evaluation Metrics Implementation Details}

\subsection{Overall Evaluation Framework}

The ONOTE benchmark uses task-specific evaluation procedures. VSU, CNC, and AST are evaluated through ground-truth matching, structural parsing, and deterministic sequence-alignment rules. SMG uses a hybrid structural protocol: deterministic checks assess
syntax, renderability, meter, measure count, and layout, while GPT-5
provides the primary 1--5 structural-compliance score. Gemini applies the
same rubric as a cross-family check. The protocol does not score
musicality, aesthetics, or creative quality. Deterministic outcomes and
judge scores are retained separately. Tables~\ref{tab:main}
and~\ref{tab:ablation} report the final GPT-5 Structural Compliance Score;
Gemini scores and tablature diagnostic subscores are not averaged into
the reported SMG value.

\subsection{Sequence Normalization}

Before scoring, notation-specific parsers normalize model outputs into
ordered tokens. Guitar string--fret pairs are converted to scientific
pitches, Jianpu degrees are resolved using key and octave marks, and staff
outputs are parsed into pitch--duration tokens. Chords are flattened in
parser order; voice and simultaneity are not retained in the final
sequence score.

\subsection{Task-Specific Evaluation Implementation Details}

\textbf{Visual Score Understanding (VSU) Evaluation}: As a classification
task, VSU parses the model-generated option identifier, normalizes its
format, and compares it exactly with the annotated ground-truth option.
It reports classification accuracy for the benchmark visual questions and
does not use semantic answer similarity or assess sequence-generation
quality.

\textbf{Cross-Format Notation Conversion (CNC) Evaluation}:
Reference and predicted outputs are parsed into ordered pitch and duration
streams. CNC uses the SequenceMatcher ratio defined in
Equation~\ref{eq:sequence-matching}, averaging pitch and duration scores
and available clefs; the guitar route compares pitches only. Unparseable
outputs receive zero.

\textbf{Audio-to-Symbolic Transcription (AST) Evaluation}:
Reference and predicted outputs are normalized into ordered
pitch--duration tokens. AST reports the matching-block precision defined
in Equation~\ref{eq:sequence-matching}; the guitar route first converts
string--fret tokens to pitches. Matching is exact after normalization,
with no onset tolerance or dynamic time alignment.

\textbf{Structurally Constrained Music Generation (SMG) Evaluation}:
SMG evaluates whether generated notation is parseable and satisfies the
declared syntax, meter, measure-count, duration, layout, and
instrument-specific constraints. ABC and Jianpu outputs are checked for
valid tokens, complete measures, and rhythmic totals. Guitar tablature is
checked for string alignment, fixed measure width, valid fret tokens, and
the disclosed 7-fret-span proxy. These checks evaluate structural
compliance only and do not measure musicality, aesthetics, creativity, or
complete performance feasibility.

\section{External Music-Knowledge Corpus}
\label{app:external-knowledge}

\subsection{Scope of the External Corpus}

The ONOTE retrieval module draws exclusively on external music-theory
materials that are independent of the benchmark data. The corpus covers
general music theory, staff notation, Jianpu, guitar tablature, pitch and
duration representation, key and harmony, standard guitar tuning,
string--fret mapping, audio-note events, metric structure, and constraints
for symbolic score generation.

\subsection{Atomic Knowledge Cards}

To reduce the retrieval of passages that are topically related but not
operationally useful, the external materials are transformed into English
atomic knowledge cards. Each card expresses one self-contained musical rule,
such as the mapping between Jianpu scale degrees and absolute pitches, the
joint interpretation of clef, staff position, and accidentals, the
relationship among guitar tuning, string, fret, and pitch, or the preservation
of simultaneity during symbolic serialization.

Each card records a unique identifier, applicable task, target notation,
rule statement, scope, conditions, exceptions, source file, page or section
locator, source hash, and licensing status.

\section{Proposition-Hypergraph Construction}
\label{app:hypergraph-construction}

\subsection{Hypergraph Representation}

ONOTE represents each validated musical rule as a proposition hyperedge
instead of decomposing it into a collection of independent pairwise
relations. Let the knowledge hypergraph be

\begin{equation}
    \mathcal{H} = (\mathcal{V}, \mathcal{E}),
\end{equation}

where $\mathcal{V}$ is the set of typed musical entities and $\mathcal{E}$ is
the set of proposition hyperedges. For proposition $p_j$, the corresponding
hyperedge is defined as

\begin{equation}
    e_j = (p_j, S_j, c_j),
\end{equation}

where $S_j \subseteq \mathcal{V}$ denotes the entities jointly participating
in the proposition, and $c_j$ stores provenance, scope, conditions,
exceptions, validation status, and version information. The incidence
relation is

\begin{equation}
B_{ij} =
\begin{cases}
1, & v_i \in S_j,\\
0, & v_i \notin S_j.
\end{cases}
\end{equation}

This representation preserves the joint semantics of multi-entity musical
rules. For example, a rule that maps a fret position to an absolute pitch
under standard guitar tuning simultaneously depends on the instrument,
tuning, string, fret, semitone offset, and resulting pitch. Decomposing this
rule into unrelated pairwise edges may discard the conditions under which
the relations jointly hold, whereas a proposition hyperedge retains the
complete rule and its source.

\subsection{Typed Entities and Disambiguation}

The hypergraph uses domain-specific entity types, including notation systems,
music symbols, pitch names, absolute pitches, scale degrees, accidentals,
enharmonic relations, onsets, durations, meters, keys, scales, chords,
instruments, strings, frets, tuning systems, measures, voices, simultaneous
events, layout structures, generation actions, and disclosed fingering constraints.

\subsection{Proposition Extraction and Validation}

External material is first divided into controlled-length textual units. A
fixed low-temperature extraction model identifies atomic propositions,
participating entities, entity types, applicability conditions, exceptions,
bilingual aliases, and provenance locators.

Every extracted proposition must be directly supported by its source card.
Propositions are rejected when they conflate unrelated rules, omit essential
conditions, introduce unsupported musical facts, or contain unresolved
entity types. The extraction output retains a bidirectional link to the
knowledge card and its original source span, allowing every retained
hyperedge to be traced back to the material from which it was derived.

\subsection{Graph Storage and Vector Indexes}

The proposition hypergraph is stored as a bipartite graph in which typed
entity nodes are connected to proposition-hyperedge nodes through incidence
relations. The graph structure is serialized in GraphML format so that each
entity, proposition, source identifier, and incidence relation remains
inspectable independently of the vector indexes.

In addition to the GraphML structure, ONOTE maintains separate storage
components for the original knowledge chunks and their metadata; the entity-vector index;the proposition-hyperedge vector index;source, alias, and provenance records; the task--notation route configuration and card registry; andan optional media-evidence vector index.

Entities and proposition hyperedges are embedded in the same
3,072-dimensional semantic space but retained in independent vector indexes.

\section{Hypergraph-RAG Adaptation and Audit}
\label{app:hypergraph-rag-audit}

\subsection{Route-Aware Atomic-Card Gating}

ONOTE defines a separate retrieval route for each task--notation combination.
The route configuration declares the task-specific and shared knowledge cards
that may be accessed by VSU, CNC, AST, and SMG under staff notation, Jianpu,
or guitar tablature. This prevents, for example, guitar-fingering rules from
being injected into an unrelated Jianpu transcription task.

After candidate generation, the adapter applies source restrictions,
route-aware filtering, atomic-rule extraction, and duplicate removal. A
candidate is retained only when its source belongs to the permitted route,
its rule can be extracted as a complete proposition, it is not marked as
benchmark-derived, and it satisfies the minimum semantic or lexical
relevance requirement. Identical or substantively duplicated rules are
merged before the evidence packet is assembled.

Route priors are part of the downstream RAG adaptation strategy rather than
an unbiased measure of graph retrieval quality. Consequently, an evaluation
of intrinsic retrieval quality should distinguish raw graph-retrieval
ranking from the final route-gated RAG ranking. Improvements produced by
pre-registered routing should not be attributed solely to the hypergraph
structure.

\subsection{Evidence-Packet Assembly and Prompt Injection}

The retained propositions are assembled under fixed item and token budgets.
Each evidence item preserves its atomic musical rule, scope or condition,
card identifier, source locator, retrieval path, and final rank. Evidence is
placed in a clearly delimited section of the model context.

The original task instruction, input content, and output grammar remain
unchanged and take precedence over retrieved text. The evidence packet may
clarify music-theory rules but cannot redefine the task, modify the requested
schema, or override benchmark constraints. Models are instructed to follow
the requested output format and not to discuss the retrieval procedure in
their final answers.

\subsection{Controlled Base--RAG Comparison}

Every Base--RAG pair fixes the evaluated model, service provider, benchmark
instance, original task prompt, output grammar, sampling parameters, maximum
output length, retry policy, parser version, and evaluator configuration.
The intended experimental difference is whether the retrieved evidence
packet is included.

Request failures, empty responses, truncation, malformed structured outputs,
and parsing failures are retained in the run record and receive the same
handling under both conditions. No task-specific repair may be activated
only for either Base or RAG.

\end{document}